\documentstyle[aps,epsf,aps]{revtex}

 \newcommand \be {\begin{equation}}
\newcommand \bea {\begin{eqnarray} \nonumber }
\newcommand \ee {\end{equation}}
\newcommand \eea {\end{eqnarray}}
 \newcommand \eps {\epsilon}
 \newcommand \bi {\bibitem}
\newcommand \s {\sigma}
\newcommand \de {\delta}

\newcommand \La {\Lambda}

\newcommand \Tr {\mbox{Tr}}
\input{epsf}
\begin{document}
\twocolumn[\hsize\textwidth\columnwidth\hsize\csname@twocolumnfalse\endcsname
\title{Classical and quantum behavior in mean-field glassy systems}
\author{Felix Ritort}
\address{Institute of Theoretical Physics\\ 
University of Amsterdam\\ Valckenierstraat 65\\ 
1018 XE Amsterdam (The Netherlands).\\ 
E-Mail: ritort@phys.uva.nl}

\date{\today}
\maketitle

\begin{abstract}
In this talk I review some recent developments which shed light on the
main connections between structural glasses and mean-field spin glass
models with a discontinuous transition. I also discuss the role of
quantum fluctuations on the dynamical instability found in mean-field
spin glasses with a discontinuous transition. In mean-field models with
pairwise interactions in a transverse field it is shown, in the
framework of the static approximation, that such instability is
suppressed at zero temperature.
\end{abstract} 
\begin{center}
 {\em Proceedings of the XIV Sitges Conference: 10-14 June 1996:
(Barcelona) Spain\\ COMPLEX BEHAVIOR IN GLASSY SYSTEMS}
\end{center}


\vfill

\twocolumn
\vskip.5pc] 
\narrowtext
\section{Introduction}
There is much
current interest in the study of disordered
systems. These are characterized by the presence of quenched disorder,
i.e. disorder which is frozen at a timescale much larger than the
typical observation time. In particular, a large amount of experimental
and theoretical work has been devoted to the study of spin
glasses\cite{BOOKS}. These are alloys where magnetic impurities are
introduced in the system. The magnetic impurities are frozen inside
the host material and the interaction between them is due to the
conduction electrons. This is the RKKY interaction which can be
ferromagnetic or antiferromagnetic depending on the distance between the
frozen impurities. Hence, the site disorder in the system leads to
frustrated exchange interactions.

But there are also intrinsically non disordered systems which display glassy
behavior as soon as they are off-equilibrium.  The most well known
examples are structural glasses \cite{GLASSES}, for instance dioxide of
silica $SiO_2$. Structural glasses are characterized by the existence of
a thermodynamic crystalline phase below the melting transition. Under
fast cooling the glass does not crystallize at the melting transition
temperature $T_M$ and it stays in the supercooled state in local
equilibrium. Instead if the temperature is slowly decreased this local
equilibrium property is lost when the cooling rate is of the same order
than the inverse of the relaxation time.

There are two main differences between glasses and spin glasses. The
first difference has been already pointed out. It is that spin glasses
are disordered systems while glasses are intrinsically clean. The
second difference emerges from experimental measurements which show
that in spin glasses there is a static quantity, the non linear
susceptibility, which diverges at a critical temperature. This implies
the existence of characteristic length scale or correlation length
which diverges at the critical point. On the contrary in real glasses
a divergence of a static susceptibility has not been observed. Hence,
while in spin glasses there is common agreement that there is a true
thermodynamic transition the situation for structural glasses is much
less clear and such a thermodynamic transition (the so called ideal
glass transition) is still only a theoretical concept.

The renewed interest in the connection between glasses and spin
glasses comes from the observation that a certain class of mean-field
microscopic spin-glass models seem to capture the essential features
which are present in structural
glasses. Furthermore, while spin glass mean-field models explicitly
contain disorder it has been recently shown that this is not an
essential ingredient and spin glass behavior is found even in
microscopic models where disorder is absent. The connection appears then
fully justified.

This connection between structural glasses and spin glasses at the
classical level has been already emphasized in other talks in this
conference (see the talks by S. Franz and G. Parisi). After reviewing
recent developments in this direction I will address a different aspect
of glassy behavior, this is the study of glassiness in the presence of
quantum fluctuations.  There are several reasons why quantum
fluctuations are interesting \cite{REVIEW_QUANTUM}.  The most compelling
reason is that the low temperature behavior of a large variety of
systems in condensed matter physics strongly depends on the presence of
disorder, for instance the quantum Hall effect and the metallic
insulator transition (see the talk by T. Kirkpatrick in these
proceedings). Another reason is more theoretical and relies on the need
for a better understanding of the role of randomness in quantum phase
transitions in the regime where there is no dissipation. The main
question we want to discuss in this proceedings concerns this last
point, i.e. how the glassy behavior which emerges in the classical
picture is modified when quantum fluctuations (i.e. non dissipative
processes) are taken into account.

The talk is organized as follows. First I will review some recent work
on the connections between glasses and spin glasses at the classical
level, putting special emphasis in the difference between spin-glass
models with continuous and discontinuous transitions. In section 3 I
will discuss the relevance of quantum fluctuations in glassy phenomena
and give a short reminder to the main results in the
Sherrington-Kirkpatrick model in a transverse field.  In section 4 I
will show that a large class of exactly solvable models with
discontinuous phase transition at finite temperature have a continuous
phase transition at zero temperature. Particular results will be
presented for the random orthogonal model. Finally I will discuss the
implications of this result and present the conclusions.

\section{Glasses v.s. spin glasses}

It has been realized quite recently that systems without disorder can
have a dynamical behavior reminiscent of spin glasses. While this
suggestive idea has to be traced back to Kirkpatrick and Thirumalai
\cite{KiTh1} only recently has it been shown how this idea works in some
microscopic models. In particular, Mezard and Bouchaud \cite{BoMe}
studied the Bernasconi model \cite{BERNA} by mapping it onto a
disordered model (the $p$-spin Ising spin glass with $p=4$) and thus
finding evidence for glassy behavior. It has also been shown in
\cite{MAPARI1} how it is possible to map the Bernasconi model (with
periodic boundary conditions) into a disordered model and solve it
exactly by means of the replica method. Within this approach it is
possible to show that both, the ordered and the disordered model, have
the same high temperature expansion. While the thermodynamics of the
models is different at low temperatures (the disordered model does not
have a crystalline state) the ordered model reproduces all the features
of the metastable glassy phase found in the disordered model including
the existence of a dynamical singularity.

There have been several microscopic models such as the sine or cosine model
\cite{MAPARI2}, fully frustrated lattices \cite{MAPARI3}, matrix models
\cite{CUKUPARI}, the Amit-Roginsky model\cite{FRHE} and mean-field
Josephson junction arrays in a magnetic field \cite{MAPARI4,CHA} (see
the talk by P. Chandra in this conference) where this approach has
been succesfully applied. The main conclusion which emerges from these
studies is that quenched disorder is not necessary to have spin glass
behavior but it can be {\em self-generated} by the dynamics. Physically
this means the following: the relaxation of the sytem becomes slower as
the temperature is decreased and the local fields, acting on the
microscopic variables of the system, can be considered as effectively
frozen. It seems also that all mean-field models where this mapping is
possible are those which show the existence of a dynamical singularity
above the static transition temperature. In fact, all models where this
equivalence has been built up are characterized by a discontinuous
transition. In the spin-glass language this corresponds to models with
one-step replica symmetry breaking transition.

In what follows I will discuss the main results concerning mean-field
spin glass models contrasting those with a continuous and discontinuous
transition. The phase transition in spin glasses is described by an
order parameter which is the Edwards-Anderson parameter (hereafter
referred as $EA$ order parameter). Because spin glasses are
intrinsically disordered systems the magnetization is not a good order
parameter since long range order is absent. In fact, below the
spin-glass transition the spins tend to freeze in certain directions
which randomly change from site to site. While spatial fluctuations of
the local magnetization are large the temporal fluctuations are quite
small and the parameter which measures the local spin glass ordering is
given by $\langle \s_i\rangle^2$.  The EA order parameter \cite{EA} is
the average of this quantity over the whole lattice,
$q_{EA}=\frac{1}{N}\sum_{i=1}^N\overline{<\s_i>^2}$. The EA parameter
varies from zero to 1. If it is very close to 1 this means that the
system is strongly frozen and thermal fluctuations are small. In
spin-glass models with a continuous transition $q_{EA}$ is zero above
the spin glass transition $T_g$ (in the paramagnetic phase) and
continuously increases as the temperature is lowered below $T_g$
(i.e. within the spin glass phase). In these models one finds
$q_{EA}\simeq (T_g-T)$ which means that the critical exponent $\beta$ is
equal to 1, a typical value found in mean-field disordered systems. The
simplest example of this class of models is the Sherrington-Kirkpatrick
model \cite{SK} defined by

\be
H=-\sum_{i<j} J_{ij}\s_i\s_j
\label{eqSK}
\ee
where the $J_{ij}$ are random Gaussian variables 
with zero mean and variance $1/N$.

In models with a discontinuous transition $q_{EA}$ is zero above $T_g$
but discontinuously jumps to a finite value below $T_g$.  Examples
of models with a transition of this type are $q$-states Potts glass
models with $q\ge 4$ \cite{GKS} and $p$-spin glass models with $p\ge 3$
\cite{GAR}. In the limit $p,q\to\infty$ both class of models converge to the
random energy model of Derrida \cite{REM} which is characterized by an
order parameter which is $1$ just below $T_g$. Because this model is a
particular limit where the energies of the configurations are randomly
distributed and also because it can be fully solved without the use of
replicas it is usually referred to as the simplest spin glass \cite{GRME}.
The two types of transitions are shown in figure 1.

Another example of a model with a discontinuous transition has been
recently introduced \cite{MAPARI2}. This is the random orthogonal model
(we will use the initials ROM in the rest of the paper to refer to this
model) defined by eq.(\ref{eqSK}) where now the $J_{ij}$ are matrix
elements of a random orthogonal ensemble of matrices,
i.e. $J_{ij}J_{jk}=\de_{ik}$. In the ROM model $q_{EA}$ jumps
discontinuously to 0.9998 at the transition point \cite{MAPARI2}. This
model can be considered as a very faithful microscopic realization of
the random energy model of Derrida. It is important to note that both
discontinuous and continuous spin glass transitions are continuous from
the thermodynamic point of view. This is related to one of the main
subtelities of spin glasses where order parameters are functions $q(x)$
in the interval [0:1] and thermodynamic quantities are integrals of
moments of these functions \cite{BOOKS}. The functional nature of the
order parameter $q(x)$ is related to the existence of an infinite number
of pure states in the spin-glass phase as shown by G. Parisi
\footnote{The local order parameter inside one state is given by the EA
order parameter which is given by the relation $q_{EA}=\max_{x} q(x)$.}. In
discontinuous transitions the $q(x)$ has a finite jump for $x\to 1$ and
all the moments $\int_0^1 q^p(x) dx$ remain continuous at the
transition, hence there is no latent heat.
\begin{figure}
\centerline{\epsfxsize=8cm\epsffile{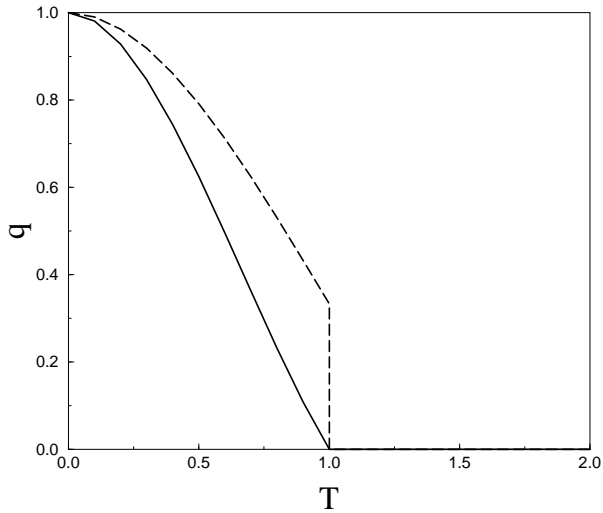}}

\caption{EA order parameter as a function of $T$ for a continuous
transition (continuous line) and a discontinuous 
transition (long dashed line).}

\end{figure}
The body of these results apply to mean-field models with long ranged
interactions. Quite surprisingly it appeared that these mean-field
models with a discontinuous transition are a nice realization of the
entropy crisis theory proposed for glasses long ago by Gibbs and
Di Marzio and later on by Adam and Gibbs in two seminal
papers\cite{AGM}. This is a heuristic theory which is based on the
Kauzmann paradox\cite{KAU} and proposes the collapse of the
configurational entropy as the mechanism for a thermodynamic glass
transition. The simplest example where this transition occurs is the
random energy model of Derrida where the phase transition coincides with
the point at which the entropy collapses to zero. This transition
corresponds to what theorists refer to as the ideal glass transition and
lies below the laboratory glass transition (see the talk by Angell in
this proceedings) which is defined as the temperature at which the
relaxation time is of order of a quite few minutes (more concretely the
viscosity is $10^{13}$ Poisse \cite{GLASSES}).

It is important to note that the ideal glass transition is a
thermodynamic transition where the configurational entropy collapses to
zero but still the total entropy can be finite since fluctuations inside
a configurational state can still be present. In the random energy
model, fluctuations in the low temperature phase are absent and the full
entropy (which gets contributions from the configurational entropy and
the entropy coming from local fluctuations inside one state) vanishes at
the glass transition $T_g$ (in what follows we will denote by $T_g$ the
ideal glass thermodynamic transition temperature).  In the ROM the
entropy at $T_g$ is of order $10^{-4}$ and can be considered quite small
(the entropy per free spin is $log(2)$).

The connection between structural glasses and spin glasses with a
discontinuous transition would not be fully acomplished if dynamics is
not taken into account. This was realized by Kirkpatrick, Thirumalai and
Wolynes \cite{KITIWO} who noted the existence of a dynamical transition
$T_d$ above the glass transition $T_g$ in connection with an instability
found in the mode coupling theory of glasses (MCT). S. Franz and
J. Hertz have shown \cite{FRHE} that the dynamical equations of the
Amit-Roginsky model (a model with pseudo-random interactions which,
nevertheless, does not contain explicit disorder) can be mapped onto the
dynamical equations of the $p$-spin spherical spin glass model with
$p=3$ \cite{CHS}. At the dynamical transition a large number of
metastable states (which grows exponentially with the size of the
system) determines an instability in the relaxational dynamics of the
system but does not induce a true thermodynamic transition. In the
region $T_g<T<T_d$ the free energy of the system is given by the
paramagnetic free energy $f_P$ but the dynamical response is fully
determined by the presence of a very large number of metastable
configurations $exp(N{\cal C}^*)$ where ${\cal C}^*$ is the so called
configurational entropy or complexity. Note that the free energy of
these metastable states is higher than the free energy of the
paramagnetic state and there is no thermodynamic transition at $T_d$. As
the temperature is decreased the number of metastable configurations
decreases and so does their free energy. When the free energy of the
metastable states equals the paramagnetic free energy there is a phase
transition. Because the number of metastable solutions with equilibrium
free energy are not exponentially large with the size of the system the
configurational entropy ${\cal C}^*$ also vanishes at this point.

The suspicious reader will find it extremely unclear how all these
quantities ($T_d$, $T_g$, $q_{EA}$, ${\cal C}^*$) can be
analytically computed. Fortunately it is not necessary to fully solve
the dynamics in order to find these quantities and there are powerful
techniques to compute them. One of the simplest procedures \cite{CWKI}
works for discontinuous transitions of the type described here and
consists in expanding the free energy around $m=1$ ($m$ parametrizes the
one step replica symmetry breaking and corresponds to the size of the
diagonal blocks with finite order parameter $q$ in the Parisi ansatz
\cite{BOOKS}).  The free energy is expanded in the following way,

\be
\beta f(q)=\beta f_P+(m-1){\cal V}(q)
\label{free}
\ee

where $f_P$ is the paramagnetic free energy (independent of $q$) and
${\cal V}(q)$ is a function called the potential
\cite{REMI}. Note in eq.(\ref{free}) that ${\cal V}(q)$ plays the role
of an entropy contribution to the paramagnetic free energy except for
the factor $(m-1)$. The dynamical transition corresponds to an
instability in the dynamics and is obtained by solving the equations,

\be \frac{\partial {\cal V}}{\partial q}=0~~~~;~~~~~~~~\frac{\partial^2
{\cal V}}{\partial q^2}=0
\label{dinamic}
\ee which yield the transition $T_d$ and the jump of the EA order
parameter $q_{EA}^d$ at the dynamical transition temperature. On the
other hand, the static transition is obtained by solving the equations

\be
\frac{\partial {\cal V}}{\partial q}=0~~~~;~~~~~~~~{\cal V}(q)=0
\label{static}
\ee and yield the transition $T_g$ and the jump of the EA order
parameter $q_{EA}^g$ at the glass transition temperature. In the range
of temperatures $T_g<T<T_d$ the complexity or configurational entropy
${\cal C}^*$ is given by the value of the potential ${\cal V}(q)$ in the
secondary minimum (see figure 2) in the region $q_{EA}^d<q<q_{EA}^g$.
For continuous transitions (such as that found in the
Sherrington-Kirkpatrick model \cite{SK}) both temperatures ($T_d$ and
$T_g$) coincide and there is no discontinuous jump of $q_{EA}$ at the
transition temperature. The behavior of ${\cal V}$ as a function of $q$
for different temperatures is shown in figure 2 for a discontinuous
transition. Note that the behavior shown in figure 2 is quite
reminiscent of a spinodal instability in first-order phase
transitions. The behavior of the potential ${\cal V}(q)$ determines the
phase transition and in particular the existence of an instability at
$T_d$. The reason why a zero mode at $T_d$ yields a divergent relaxation
time in the dynamics is related to the one of the most prominent
features of glassy systems: the dominance of an exponentially large of
metastable states ($\exp(N{\cal C}^*)$) at that
temperature\cite{KITIWO,THEO}. Note that in mean-field theory metastable
states have an infinite lifetime, the time to jump from the
metastable glassy phase to the paramagnetic state being equal to
$exp(N{\cal B}^*)$ where ${\cal B}^*=\max_{0<q<q^g_{EA}}{\cal V}(q)$ is
the height of the free energy barrier which separates the metastable and
the paramagnetic phase. The extension of this approach to the
computation of thermodynamic quantities below $T_d$ in the metastable
glassy phase has been considered in \cite{REMI}.

Summarizing, mean field spin-glass models with a discontinuous
transition are good models to describe real glasses. The role of
disorder is not essential and can be {\em self-generated} by the
dynamics.  These models show a thermodynamic transition (the ideal glass
transition $T_g$) where the configurational entropy ${\cal C}^*$ collapses
to zero and the EA order parameter jumps to a finite value. Concerning
dynamics, these models are described by the mode coupling equations
which are a good description of relaxational processes in real glasses
at temperatures above but not too close to $T_d$. The instability found
at $T_d$ is a mean-field artifact which should be wiped out by including
activated processes over finite energy barriers. In this sense, mode
coupling equations are genuine mean-field dynamical equations. The
interested reader can find more details in \cite{MEZARD}.
\begin{figure}
\centerline{\epsfxsize=8cm\epsffile{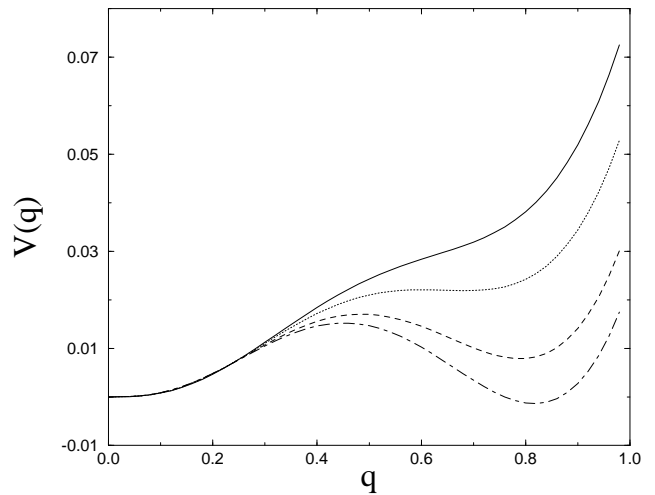}}

\caption{Potential ${\cal V}(q)$ for a spin glass model with a discontinuous
transition. The different regimes are: $T>T_d$ (continuous line),
$T=T_d$ (dotted line), $T_g<T<T_d$ with complexity ${\cal
C}^*=\min_{q>0}{\cal V}(q)$ (dashed line), $T=T_g$
with ${\cal C}^*={\cal V}(q_{EA}^g)=0$ (dot-dashed line)}
\end{figure}

\section{Ising spin glasses in a transverse field}

There is much recent interest in the study of quantum phase transitions
(see the talks by T. Kirkpatrick, R. Oppermann and H. Rieger in this
proceedings). These transitions appear at zero temperature when an
external parameter is varied. For a certain critical value of this
parameter the system enters into the disordered phase.  The
general problem can be put in the following way. Let us consider the
following Hamiltonian,

\be
{\cal H}={\cal H}_0+ \epsilon P
\label{eq3a}
\ee

where ${\cal H}_0$ stands for the unperturbed Hamiltonian and $P$ is a
perturbation which does not commute with ${\cal H}_0$, i.e. $[{\cal
H}_0,P]\ne 0$ and $\epsilon$ denotes the strength of the
perturbation. Let us suppose that the system for $\epsilon=0$ is in
an ordered phase at $T=0$. As the control parameter
$\eps$ is varied and the strength of the perturbation $P$ increases, the
new ground state of ${\cal H}$ becomes a mixture of all the eigenstates
of ${\cal H}_0$ and this tends to disorder the system. For a certain
value of the control parameter $\eps$ the systems fully disorders. The
effect of the control parameter $\eps$ in quantum phase transitions is
rather similiar to the effect of temperature in classical
systems. The difference is that now quantum effects are non dissipative
while thermal effects are. 

In the realm of disordered systems it is essential to understand the
role of disorder in quantum phase transitions. The rest of this talk
will be devoted to discuss this problem in the framework of disordered
mean-field spin glass models as presented in the previous section.  Let
us discuss how the classical glassy scenario is modified in the presence
of quantum fluctuations. This problem has been already addressed in the
literature, in particular the question wether replica symmetry breaking
survives to the effect of quantum fluctuations. Large amount of work have
been devoted to the study of the Sherrington-Kirkpatrick Ising spin
glass in a transverse field \cite{BM,USA,ISYA,GOLA,KITHLI}. The model is
defined by,

\be {\cal H}={\cal H}_0+\Gamma
P=-\sum_{i<j}J_{ij}\s_i^z\s_j^z-\Gamma\sum_i\s_i^x
\label{eq1}
\ee

where $\s_i^z,\s_i^x$ are the Pauli spin matrices and $\Gamma$ is the
the transverse field which plays the role of a perturbation.  The
indices $i,j$ run from 1 to $N$ where $N$ is the number of sites. The
$J_{ij}$ are random variables Gaussian distributed with zero mean and
$1/N$ variance. Note that for $\Gamma=0$ this model reduces to the
classical Sherrington-Kirkpatrick model which has a continuous
transition to a replica broken phase with local spin-glass ordering in
the $z$ direction. The effect of the transverse field is to mix 
configurations and to supress the order in the $z$ direction. For a critical
value of $\Gamma$ the ordering in the $z$ direction is completely
suppressed at the expense of ordering in the $x$ direction. Note that the
effect of a perturbation in the $z$ direction in the form of a
longitudinal magnetic field $P=-\sum_i\s_i^z$ has a quite different
effect on the phase transition.  The reason is that it commutes with
the unperturbed Hamiltonian, hence it does not mix configurations.

The phase diagram of the model eq.(\ref{eq1}) is reproduced in figure
3. There are two phases, the quantum paramagnetic (QP) and the quantum
glass phase (QG). The transition is continuous all along the phase
boundary and the QG phase resembles the classical one where
replica symmetry is broken and a large number of pure states, with
essentially the same free energy, contribute to the thermodynamics
\footnote{The extensive free energies of the different solutions only
differ by finite - i.e. non extensive- quantities.}. A large body of
information on the SK model with a transverse field has been obtained
from spin summation techniques \cite{GOLA} and perturbative expansions
\cite{ISYA}. But only very recently has a full understanding of this
model been achieved through seminal work by Miller and Huse\cite{MIHU} and
in an independent way by Ye, Read and Sachdev \cite{YESARE}. They have
been able to obtain the frequency response of the system as well as the
crossover lines which separate regimes where thermal or quantum
fluctuations are dominant. The values of the critical exponents as well
as the nature of finite size corrections in the quantum critical point
have been also numerically checked in \cite{LR}.

\begin{figure}
\centerline{\epsfxsize=8cm\epsffile{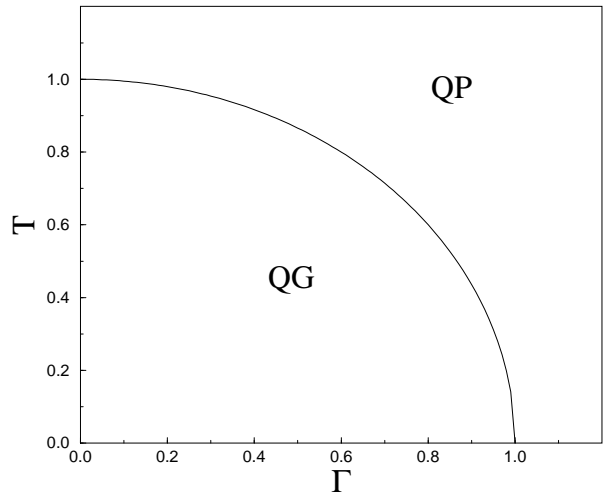}}

\caption{Phase diagram for the SK model in a transverse field.  The
critical boundary separates the quantum paramagnetic
phase (QP) from the quantum glass phase (QG).}
\end{figure}

In this framework one would like to understand the role of quantum
fluctuations in systems with a discontinuous transition.  In particular
it is relevant to understand how quantum fluctuations could modify the
dynamical instability $T_d$ obtained in classical systems within the
mode coupling approach. Note that at zero temperature the entropy must
vanish and the idea of an entropy crisis as the transverse field
$\Gamma$ is varied is nonsense. On the other hand, since the statics and
dynamics are inextricably linked in quantum phase transitions it is
interesting to ask to what extent a dynamical instability (which is not
related to any static singularity) can survive at zero temperature.

In this proceedings I want to discuss some recent results in a general
family of solvable models which strongly suggest that the dynamical
instability predicted in the mode coupling approach is completely
suppressed at zero temperature when quantum fluctuations are taken
dominant\cite{RIT}.

In particular we will show, always within this class of models, that the
discontinuous transition becomes continuous at zero temperature.
Particular results will be shown for the ROM (random ortoghonal
model). The implications of this and other results will be also
discussed.

\subsection{Mean-field models with pairwise interactions}

The family of exactly solvable models we are interested in are quantum
Ising spin glasses with pairwise interactions in the presence of
a transverse field. These are described by the Hamiltonian,

\be
{\cal H}=-\sum_{i<j}J_{ij}\s_i^z\s_j^z-\Gamma\sum_i\s_i^x
\label{eq4a}
\ee

where $\s_i^z,\s_i^x$ are the Pauli spin matrices and $\Gamma$ is the
the transverse field. The indices $i,j$ run from 1 to $N$ where $N$ is
the number of sites. The $J_{ij}$ are the couplings taken from an
ensemble of random symmetric matrices. 

Details about how to analytically solve the quantum model (\ref{eq4a})
can be found in \cite{RIT}. This are based on matrix theory
techniques introduced in \cite{MAPARI1} to solve glassy models without
disorder.  Here we want we present some of the main results of the
analysis of \cite{RIT}. In practice the simplest way to solve the
model eq.(\ref{eq4a}) is by means of the replica trick where we compute
the average over the disorder of the $n$-th power of the partition
function making the analytical continuation $n\to 0$ at the end,

\be
\beta f=\lim_{n\to 0} \frac{\overline{Z_J^n}-1}{n}
\label{eq4b}
\ee

where
\be \overline{Z_J^n}=\int [dJ]
\Tr\, exp(\sum_{a=1}^n\,{\cal H}^a)
\label{eq4c}
\ee 

and $\int [dJ]$ means integration over the random ensemble of
matrices. This integral can be done using known methods in matrix
theory \cite{ITZU,MAPARI2}. The final result of eq.(\ref{eq4c}) can be
written in terms of a generating function $G(x)$ which depends on the
particular ensemble of $J_{ij}$ couplings via its spectrum of
eigenvalues. For the two examples we will consider here we
have $G_{SK}(x)=\frac{x^2}{2}$ ($SK$ model) and
$G_{ROM}(x)=\frac{1}{2}
log(\frac{\sqrt{1+4x^2}-1}{2x^2})+\frac{1}{2}\sqrt{1+4x^2}-
\frac{1}{2}$ (ROM model). 

By going to imaginary time and using the Trotter-Suzuki
breakup we end up with a
closed expression for the free energy. The final result is,

\be 
\overline{Z_J^n}=\int dQ\,d\La exp(-N F(Q,\La))
\label{eq4d}
\ee
where 
\be
F(Q,\La)=-\frac{nC}{N}+\frac{1}{M^2}\Tr(Q\La)-\frac{1}{2}\,\Tr G(AQ)-\log(H(\La))
\label{eq4e}
\ee where the constants $A$, $B$ and $C$ are given by
$A=\frac{\beta}{M}; B=\frac{1}{2}\log(coth(\frac{\beta\Gamma}{M}));
C=\frac{MN}{2}\log(\frac{1}{2} sinh(\frac{2\beta\Gamma}{M}))$.  The
order parameters now depend on two set of indices: the replica index and
the time indices corresponding to the imaginary time direction. The time
indices $t,t'$ go from $1$ to $M$ where $M$ is the length of the
discretized imaginary time direction. The order parameters are
$Q_{ab}^{tt'},\La_{ab}^{tt'}$ (the $\La$ have been introduced as
Lagrange multipliers in the saddle point equations) and the trace $\Tr$
is done over the replica and time indices.  The term $H(\La)$ is given
by, \be H(\La)=\sum_{\s}\,exp(\sum_{ab}\frac{1}{M^2}\sum_{t
t'}\La_{ab}^{t t'}\s_a^t\s_b^{t'}\,+\,B\sum_{at}\s_a^t\s_a^{t+1})
\label{eq4f}
\ee and the free energy is obtained by making the analytic continuation
$\beta f=\lim_{n\to 0} \frac{F(Q^*\La^*)}{n}$ where $Q^*,\La^*$ are
solutions of the saddle point equations,
$\La_{ab}^{tt'}=\frac{AM^2}{2}\Bigl (G'(AQ)\Bigr )^{tt'}_{ab}$ and
$Q_{ab}^{tt'}=\langle \s_a^t\s_b^{t'}\rangle$. The average
$\langle(\cdot)\rangle$ is done over the effective Hamiltonian in
(\ref{eq4f}). For $a=b$ we have translational time invariance and the
order parameter becomes independent of the replica index, i.e 
$Q_{aa}^{tt'}=R(|t-t'|)$ 

Once we have written a closed expression for the free energy
eq.(\ref{eq4e}) one can obtain the static and dynamical transition
temperatures according to eq.(\ref{dinamic},\ref{static}). Such a
solution always exists for models with a single quantum paramagnetic
phase. In particular, using eq.(\ref{dinamic}), a closed expression for
the dynamical instability can be obtained (see \cite{RIT}). For a
continuous transition this equation can be written in the simple form

\be \chi_0^2
G''(\chi_0)=1
\label{eq4g}
\ee where $\chi_0=\beta\hat{R}_0$ is the longitudinal magnetic
susceptibility and $\hat{R}_p=M^{-1}\sum_{t=0}^{M-1}e^{i\omega_p t}
R_t$ is the Fourier transformed order parameter $R(t)$ in terms of the
Matsubara frequencies $\omega_p=\frac{2\pi p}{M}$.

Our main aim is to compute the order of the transition. We already know
that some models within the family eq.(\ref{eq4a}) (for instance, the
ROM) have a classical discontinuous transition with a dynamical
instability above the static transition. Is the discontinuous nature of
this transition changed in the presence of quantum fluctuations?

To answer this question we consider the static approximation introduced
by Bray and Moore in the context of quantum spin glasses\cite{BM}. This
approximation considers $R(t-t')$ to be constant which amounts to take
into account only the zero frequency behavior $p=0$ (small energy
fluctuations) in the set of order parameters $\hat{R}_p$. We will later
comment on the validity of this approximation.

Using this approximation one can write closed expressions for the
paramagnetic free energy $f_P$ and the complexity $C$. It is found
\cite{RIT} that at $T=0$ the dynamical transition $T_d$ and the static
transition $T_g$ coincide and the complexity vanishes. The transition
then becomes continuous. The value of the critical field and all the
thermodynamic observables at the critical point at zero temperature can
be expressed in terms of the longitudinal susceptibility $\chi_0$ which
satisfies the following simple set of equations,

\be
\chi_0^2 G''(\chi_0)=1~~;~~\Gamma-\frac{1}{\chi_0}=G'(\chi_0);
\label{eq5g}
\ee 

Note that the first of eq.(\ref{eq5g}) can be obtained derivating the
second of eq.(\ref{eq5g})respect to $\chi_0$.  At the critical point
the internal energy is given by $U=-\Gamma_c$ and the entropy
$S=\frac{1}{2}(G(\chi_0)+1-\Gamma_c\chi_0+log(\Gamma_c\chi_0))$.  Above
the critical point, always at zero temperature, the second of
eq.(\ref{eq5g}) is still valid and yields the susceptibility as a
function of $\Gamma$.

To go beyond the static approximation we should consider all the
Matsubara modes $\hat{R}_p$ in the saddle point equations. The
difficulty of this problem is similar to that found in strongly
correlated systems where an infinite set of parameters has to be computed in
a self-consistent way\cite{KR}. Nevertheless we expect the order of the
transition to be correctly predicted. The essential idea is that for a
continuous transition at zero temperature the gap vanishes. It would be
quite surprising that higher frequency modes they could
drastically modify the low frequency behavior. The order of the
transition should not be determined by the decay of the correlation
$R_t$ in imaginary time but for its infinite time limit which is the
EA parameter at the transition point \cite{YESARE}.

In the next subsection we analyze our results for the particular case of
the ROM model and compare them with those obtained in case of the SK
model.

\subsection{Results for the ROM}

As has been already said, the ROM has a classical discontinuous
transition at zero transverse field where $q_{EA}^d\simeq 0.962$ at the
dynamical transition $T_d\simeq 0.134$ and $q_{EA}^d\simeq 0.9998$ at
the static transition $T_g\simeq 0.065$. At the static transition the
configurational entropy or complexity ${\cal C}^*$ vanishes and the total
entropy is of order of $10^{-4}$. The locations of the static and
dynamical transitions can be evaluated within the static approximation
to obtain the results shown in figure 4. Both transition temperatures
decrease as a function of the transverse field merging
into the same point at zero temperature as one would expect for a continuous
transition.  In figure 5 we show the EA order parameter
$q=\overline{<\s^z>^2}$ as a function of $\Gamma$ as we move along the
static ($q^g_{EA}$) and dynamical ($q^d_{EA}$) phase boundaries. Note
that both EA order parameters $q^d_{EA}$ and $q^g_{EA}$ vanish at zero
temperature like $T^{\frac{1}{2}}$, hence the jump in the order
parameter dissapears at zero temperature.

\begin{figure}
\hfill{\epsfxsize=7cm\epsffile{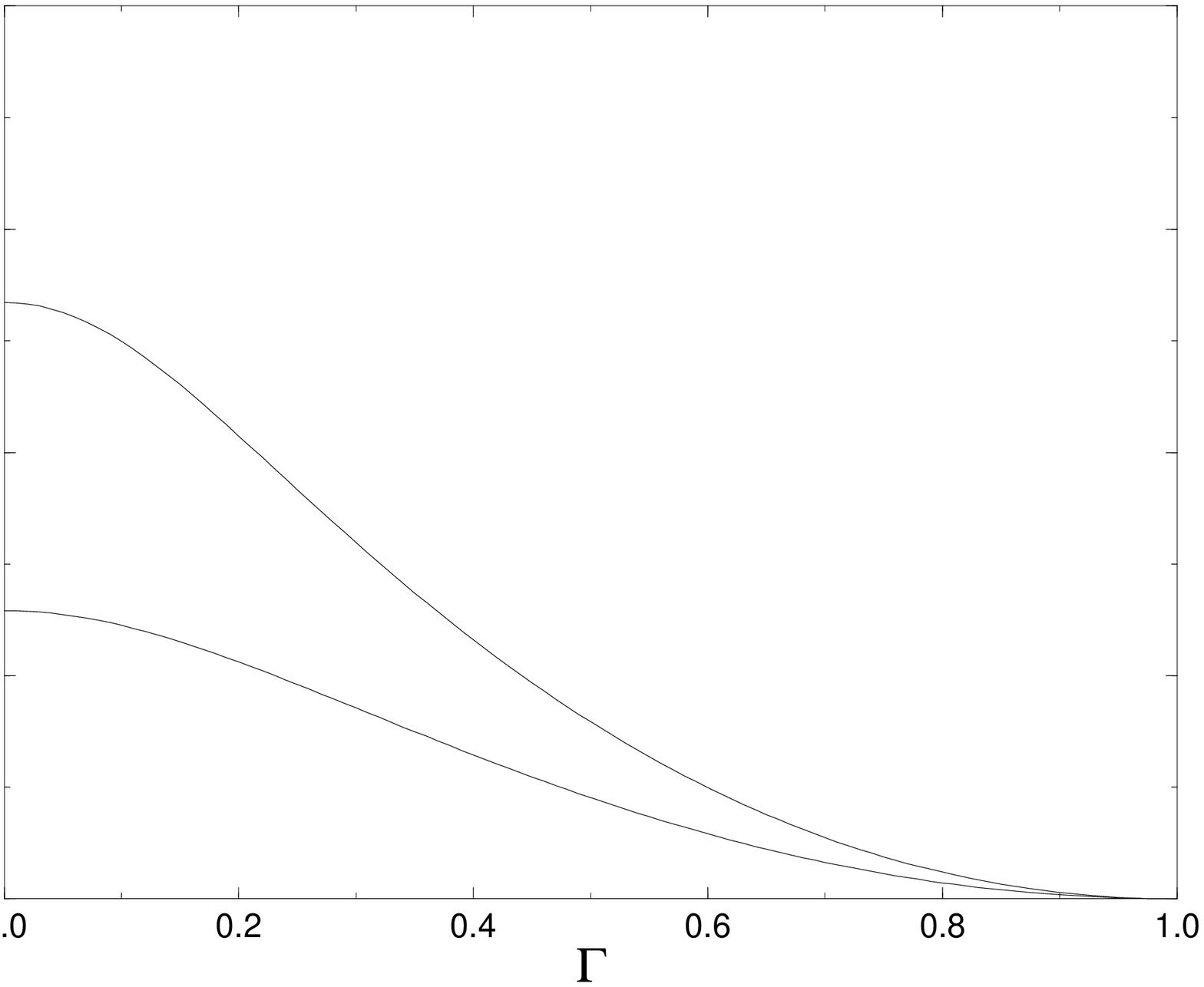}}\hfill
\caption{Phase boundaries $T_g(\Gamma)$ (lower line) and
$T_d(\Gamma)$ (upper line) in the ROM in the static approximation.  At
zero transverse field $T_g\simeq 0.0646,T_d\simeq 0.1336$.}
\end{figure}

By substituting the particular function $G(x)$ for the ROM model in the
second of eq.(\ref{eq5g}) the susceptibility at zero temperature in the
QP phase can be analytically obtained in the static approximation. One
finds $\chi_0^{ROM}=\frac{\Gamma}{\Gamma^2-1}$ which diverges at the
critical field $\Gamma_c=1$. This is quite different to what is found in
the SK model where $\chi_0^{SK}=\frac{\Gamma-\sqrt{\Gamma^2-4}}{2}$ and
is finite at the critical point $\Gamma_c=2$. Note that in the static
approximation the critical field is given by the maximum eigenvalue of
the coupling matrix $J_{ij}$.

\begin{figure}
\centerline{\epsfxsize=7cm\epsffile{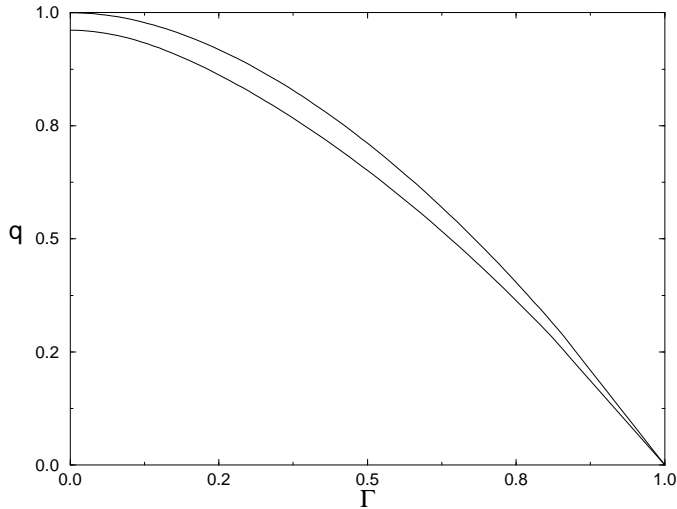}}

\caption{EA order parameter $q_{EA}^g$ (upper line) and $q_{EA}^d$
(lower line) in the ROM on the static and dynamical phase boundaries
boundaries as a function of the transverse field. At zero transverse
field $q_{EA}^g\simeq 0.99983,q_{EA}^d\simeq 0.961$. Both $q_{EA}^g$
and $q_{EA}^d$ vanish linearly with $T^{\frac{1}{2}}$ at zero
temperature.}
\end{figure}

Unfortunately, as we have said before, the static approximation gives
incorrect results for the thermodynamic quantities. In particular, the
entropy is finite at zero temperature in the SK model \cite{KITHLI} and
infinite in the ROM case. It is important to note that despite these
failures of the static approximation some exact results can still be
derived, in particular from eq.(\ref{eq4g}).  When the transition is
continuous equation (\ref{eq4g}) is exact. One finds for the ROM model
that the longitudinal susceptibility $\chi_0$ diverges at the critical
point. This is different to what happens in the SK model where
$\chi_0=1$ at the critical field. To see clearly the general
implications of this result for continuous quantum phase transitions we
observe that $\chi_0$ is given by the decay in imaginary time of the
correlation function $R(t)$ via the relation, $\chi_0=\int_0^{\beta}
R(t)dt$ (now the time $t$ has become a continuous variable in the
$M\to\infty$ limit). In the SK model the large time behavior of the
$R(t)$ has been obtained \cite{MIHU,YESARE}. It is found that
$R(t)\simeq t^{-2}$ which at zero temperature yields a finite value of
the susceptibility. This decay is necessarily slower in the case of the
ROM model where $\chi_0=\infty$. Because the decay of $R(t)$ in
imaginary time is related to the quantum critical exponents via the
relation $R(t)\simeq t^{-\frac{\beta}{z\nu}}$ the $z$ exponent is
probably not a universal quantity and different mean-field models with
continuous transitions may have different exponents. For the SK model
$z=2,\beta=1,\nu=\frac{1}{4}$ yield the correct decay. These results
suggest that the value of the exponent $z$ is larger than $2$ in the
ROM.

The exact computation of the critical exponents and the full analysis
of the problem beyond the static approximation in the ROM remains an
interesting open problem.

\section{Conclusions}

In this talk I have reviewed some recent developments in the theory of
spin glasses which allow for a comparison between glasses and spin glass
models with a discontinuous transition. I have stressed that the two
most common approaches to the glass transition, the thermodynamic
approach based on the Adam-Gibbs theory and the dynamical approach based
on mode coupling theory, appear quite naturally in the framework of
mean-field spin glass models with a discontinuous transition. I have
also stressed that disorder is not necessary to find spin-glass behavior
and a large family of non disordered mean-field models indeed show
spin-glass behavior.  The connection between disordered spin glasses and
structural glasses, at least at the mean-field level, appears to be
fully justified. Going beyond the mean-field level is a major open
problem.  It is well accepted that the dynamical instability at $T_d$ is
an artifact of the mean-field theory but it is unclear if the
entropy crisis survives in finite dimensions (see the talk of S. Franz
in these proceedings).

We have also discussed the role of quantum fluctuations in glassy
systems by studying the Ising spin glass in a transverse field. In
quantum phase transitions statics and dynamics are inextricably
linked. Then it is of relevance to understand the role of complexity in
non relaxational quantum dynamics.  We have shown that the classical
glassy scenario with a dynamical transition above the thermodynamic
transition is modified in the presence of quantum fluctuations. This
result has been obtained in the framework of models with two spin
interactions in the presence of a transverse field. In models with a
discontinuous finite temperature transition we have shown, using the
static approximation, that the transition becomes continuous at $T=0$
and there is no room for a metastable glassy phase. We have argued in
favour of this result even beyond the static approximation. Particular
results have been presented for the ROM model where it has been shown
that some critical exponents at the quantum phase transition should
differ from the mean-field exponents derived in the case of the SK
model. It is still too soon to understand the implications of this
result which deserves further investigation. The removal of the
instability at $T_d$, even in mean-field theory, could be a general
consequence of the non dissipative nature of quantum processes.  How
general this result is for other type of models remains an
interesting open problem. In this direction it would be very instructive
to address the problem presented here within the approach developed in
\cite{MIHU,YESARE} for the SK model as well as taking this research
further by studying the zero temperature dynamical transition in quantum
$p$-spin glass models \cite{GOLDS} and Potts glass models \cite{SE}.

{\bf Acknowledgements.} I am very grateful to the following colleagues
for fruitful collaborations in this and related subjects: J. V. Alvarez,
S. Franz, D. Lancaster, E. Marinari, Th. M. Nieuwenhuizen, F. G. Padilla
and G. Parisi.  I acknowledge to the Foundation for Fundamental Research
of Matter (FOM) in The Netherlands for financial support through
contract number FOM-67596.

\end{document}